\documentstyle[psfig,natbib209]{mn} 

\def\ifundefined#1{\expandafter\ifx\csname#1\endcsname\relax}

\def\la{\mathrel{\hbox{\rlap{\hbox{\lower4pt\hbox{$\sim$}}}\hbox{$<$}}}}
\def\ga{\mathrel{\hbox{\rlap{\hbox{\lower4pt\hbox{$\sim$}}}\hbox{$>$}}}}

\newcommand{\be}{\begin{eqnarray}}
\newcommand{\ee}{\end{eqnarray}}

\ifundefined{nuc}\def\nuc#1#2{\relax%
\ifmmode{}^{#1}{\protect\text{#2}}\else${}^{#1}$#2\fi}\else\relax\fi

\newcommand{\etal}{et al.}
\newcommand{\gcm}{g~cm$^{-3}$}
\newcommand{\kmps}{km~s$^{-1}$}

\newcommand{\msol}{\ifmmode{{\rm M}_\odot}\else{M$_\odot$}\fi}
\newcommand{\foe}{\ifmmode{10^{51}}\else{$10^{51}$}\fi}
\newcommand{\nni}{\nuc{56}{Ni}}
\newcommand{\xni}{\ifmmode{{\rm X}_{\rm Ni}}\else{X$_{\rm Ni}$}\fi}
\def\ang{\hbox{\AA}}
\def\Teff{\ifmmode{T_{\rm eff}}\else{\hbox{$T_{\rm eff}$} }\fi}
\def\Rzero{\ifmmode{R_0}\else{\hbox{$R_0$} }\fi}
\newcommand{\vno}{\ifmmode{v_0}\else{\hbox{$v_0$} }\fi}

\newcommand{\heoh}{\ifmmode{{\rm [He/H]}}\else{[He/H]}\fi}
\newcommand{\heoo}{\ifmmode{{\rm He/O}}\else{He/O}\fi}
\newcommand{\coo}{\ifmmode{{\rm C/O}}\else{C/O}\fi}
\newcommand{\mnras}{MNRAS}

\begin{document}

\bibliographystyle{natbib-apj}

\title{Preliminary Spectral Analysis of SN 1994I}

\author[E.~Baron et al.]{E.~Baron$^{1}$, P.~H.~Hauschildt$^{2}$,
D.~Branch$^{1}$, R.~P.~Kirshner$^{3}$, and~A.~V.~Filippenko$^{4}$\\
$^{1}${Dept. of Physics and Astronomy, University of Oklahoma, 440 W.
Brooks, Rm 131, Norman, OK 73019-0225.}\\ $^{2}${Dept. of Physics and
Astronomy, Arizona State University, Tempe, AZ 85287-1504.}\\
$^{3}${CfA, 60 Garden Street, Cambridge, MA 02138.}\\ $^{4}${Dept. of
Astronomy, University of California, Berkeley, CA 94720-3411.}}

\maketitle

\begin{abstract}
We present optical spectra of the Type Ic supernova 1994I in M51 and
preliminary non-LTE analysis of the spectra.  Our models are
not inconsistent with the explosions of C+O cores of massive stars.  While
we find no direct evidence for helium in the optical spectra, our
models cannot rule out small amounts of helium. More than 0.1~\msol\
of helium seems unlikely.
\end{abstract}
\begin{keywords}
{stars: evolution --- supernovae: general --- supernovae:
individual (SN 1994I). } 
\end{keywords}

\section{Introduction}
The last few years have produced well observed supernovae of nearly
every subclass. Of all the subclasses, the progenitors of Type Ib/c
supernovae are the least well understood. Type I supernovae are
classified by the lack of strong Balmer lines in their spectra.
Whereas Type Ia supernovae are characterized by strong Si~II
lines and Type Ib supernovae by prominent He~I
lines, in Type Ic supernovae these features are weak or absent
altogether. Supernova 1994I in M51 has been classified as a SN Ic
\cite[]{phill94i_iau,clocc94i_iau}. 
While there is little doubt that SNe Ia are the result of the complete
disruption of a white dwarf, several models have been proposed for the
progenitors of Type Ib and Type Ic supernovae. The prominence of
He~I lines in Type Ib supernovae suggests that these SNe are due
to the collapse of the helium core of a massive star, whose hydrogen
envelope has been lost in a wind, or through a binary
interaction. However, such a model fails to fit the long time behavior of the
light curve of SN 1984L \cite[]{byb93}. Several groups have suggested
that the progenitors of SNe Ic are due to collapse of C+O cores of
massive stars, where both the hydrogen and helium envelopes have been
stripped in either a Wolf-Rayet phase, or via binary interaction
\cite[]{wh90,fps90,pod92,wlw93,nomnat94i}. Based on the very early light
curve of SN 1994I,  \cite{nomnat94i} have
proposed that single star Wolf-Rayet models are ruled out and have
described several possible evolutionary paths for the progenitor of SN
1994I.  Further modeling of the light curve has suggested a very low
mass for the ejecta, $M_{\rm ejecta} \approx 0.9 - 1.4$~\msol\ 
\cite[]{iwsn94i94,wlw95,ybb95}. 

\section{Spectral Models}

In order to provide model constraints 
the observed spectra must be modeled  in detail. We use the
generalized stellar atmosphere code {\tt PHOENIX 4.8} to compute 
model atmospheres and synthetic spectra for SN 1994I. This is an
updated version of the code used for the analysis of the early spectra
of Nova Cygni 1992 \cite[]{hsawss94}, described there, in
\cite{h2o94}, in \cite{phhnov95}, 
in  \cite{b93j2}, and in \cite{b93j3}.  Thus, here we only briefly outline
the calculational methods.

{\tt PHOENIX} uses an accelerated $\Lambda$-iteration (ALI or operator
splitting) method to solve the time independent, spherically
symmetric, {\sl fully} relativistic radiative transfer equation for
lines and continua, to all orders in $v/c$ including the effects of
relativistic Doppler shift, advection, and aberration
\cite[]{phhs392}.  The multi-level, non-LTE rate equations are solved
self-consistently using an ALI
method \cite[]{rh91,phhcas93}. Models atoms  for H~I, He~I, He~II,
Mg~II, Ca~II, Na~I, and Ne~I have been
implemented. Simultaneously we solve for the
special relativistic condition of radiative equilibrium in the
Lagrangian frame \cite[]{phhre92} using either a partial linearization
or a modified Uns\"old-Lucy temperature correction scheme.  The
relativistic effects, in particular the first order effects of
advection and aberration, are important at the high expansion
velocities observed in typical SNe \cite[]{hbw91}.

In addition to the non-LTE lines, the models include,
self-consistently, line blanketing of the most important ($\approx
10^5$) metal lines selected from the latest atomic and ionic line list
of \cite{kurucz93}.  The entire list contains close to
42 million lines, but not all of them are important for the case at
hand; the lines are dynamically selected according to whether the
ratio $\Gamma \equiv \kappa_l / \kappa_c $ is larger than a
pre-specified value (usually $10^{-3}$), where $\kappa_l$, $\kappa_c$
are the absorption coefficients in the line center, and in the
corresponding (LTE $+$ non-LTE) continuum, respectively. In the
subsequent radiative transfer calculations all lines selected in this
way are taken into account as individual lines by opacity sampling,
and all others from the large line list are neglected. We treat line
scattering in the metal lines by parameterizing the albedo for single
scattering, $\alpha$. The calculation of $\alpha$ would require a full
non-LTE treatment of {\sl all} lines and continua, which is outside of
the scope of this paper.  Tests have shown that as a direct result of
the velocity gradient in nova and supernova photospheres, the shape of
the lines does not depend sensitively on $\alpha$ and our approach is
a reasonable first approximation. Therefore, we adopt a constant value
of $\alpha=0.95$ for all metal lines.  The continuous absorption and
scattering coefficients are calculated using the cross-sections as
described in \cite{hwss92}.

The effects of non-thermal collisional ionization by primary electrons
produced by collisions with gamma rays due to the decay of $^{56}$Ni
and $^{56}$Co are modeled using the continuous slowing down
approximation \cite[]{gg76,swartz91}. We have also included
ionizations due to secondary electrons \cite[]{swartz91}, but most of
their energy is thermalized and therefore has only a small effect on
the level populations \cite[]{meyerott80}. The collisional cross
sections are taken from the work of 
\cite{lotz67a,lotz67b,lotz68a,lotz68b,lotz68c}.

The model atmospheres are characterized by the following parameters
\cite[see][for details]{hwss92,b93j3}: (i) the reference radius
$\Rzero$, which is the radius where the continuum optical depth in
extinction at 5000~$\ang$ ($\tau_{\rm std}$) is unity, (ii) the
effective temperature \Teff\unskip, which is defined by means of the
luminosity, $L$, and the reference radius, $\Rzero$, ($\Teff=(L / 4\pi
\Rzero^2 \sigma )^{1/4}$ where $\sigma$ is Stefan's constant), (iii)
the density structure parameter, $v_e$, ($\rho(r)\propto
\exp(-v/v_e)$), or $N$ ($\rho(r)\propto (v/\vno)^{-N}$), (iv) the
expansion velocity, $\vno$, at the reference radius, (v) the density,
$\rho_{\rm out}$, at the outer edge of the envelope, (vi) the albedo
for line scattering (metal lines only, here set to $0.95$), (vii) the
statistical velocity $\xi$, treated as depth-independent isotropic
turbulence and set to 50~km~s$^{-1}$ for the models presented here,
and (viii) the element abundances.

In order to minimize the parameters in setting the element abundances
we begin with a solar abundance \cite[]{ag89}, and then ``burn''
hydrogen to helium with a final number ratio given by $\heoh =
\log_{10}(n_{He}/n_{H})$. Next we  ``burn'' helium to carbon and oxygen
with a final number ratio \heoo, and a production ratio (by number) of
carbon to oxygen \coo. Finally we allow metals heavier than oxygen to
be scaled by the factor $z$. Thus, our abundance parameters consist
of \heoh, \heoo, \coo, and $z$. In order to further reduce the size
of parameter space we have used the results from the models of 
\cite{iwsn94i94} and \cite{wlw95} as
guidance to 
fix the parameter \coo = 0.5.
The mass fractions for the models presented are listed in Table~\ref{tab1}.

\begin{table}
\caption{Abundances of Presented Models\label{tab1}}
\medskip
\begin{tabular}{lccc}
Model&$X_{He}$&$X_{C}$&$X_{O}$\\
April 3, low helium&$0.05$&$0.25$&$0.68$\\
April 3, high helium&$0.35$&$0.17$&$0.46$\\
April 11&$1.8\times 10^{-4}$&$0.27$&$0.71$\\
April 18&$1.8\times 10^{-4}$&$0.27$&$0.71$\\
\end{tabular}

\medskip
The mass fractions of helium, carbon, and oxygen used in the models.
\end{table}

For convenience, we have assumed an explosion date of March 30, 1994
to scale the radii of our models. We have confirmed that changes of a
few days in the explosion date have very small effects on the shape of
the output spectrum although the absolute flux obviously scales with
the square of the radius, which we account for.  We adopt an
extinction to the supernova of $E(B-V) = 0.45$~mag, which we discuss
further in \S{4}. In order to avoid calculational difficulties, we have
treated He~II in LTE.

Since our models do not include NLTE treatments of all the relevant
ions (in particular O~I, O~II, C~I, C~II,
Si~II, and Fe~II are treated in LTE), we cannot, as yet,
provide quantitative abundance estimates.  We can, however, draw
qualitative conclusions based on our modeling.  In future work we will
present calculations including
 a $\approx 617$ level Fe~II model atom \cite[]{phheb95}. 

While we only present results of our ``best fits'' (as determined by
eye, comparing to the complete observed spectrum) we have in fact
calculated more than 100 models, where we have varied the He abundance
by factors of 10,000, we have examined models from 1/2 solar to 6
times solar metallicity, and we have varied the C/O production ratio
by a factor of 3. All of the models discussed in this paper are
available upon request.

\section{Spectra and Results}

\begin{figure*}
\psfig{file=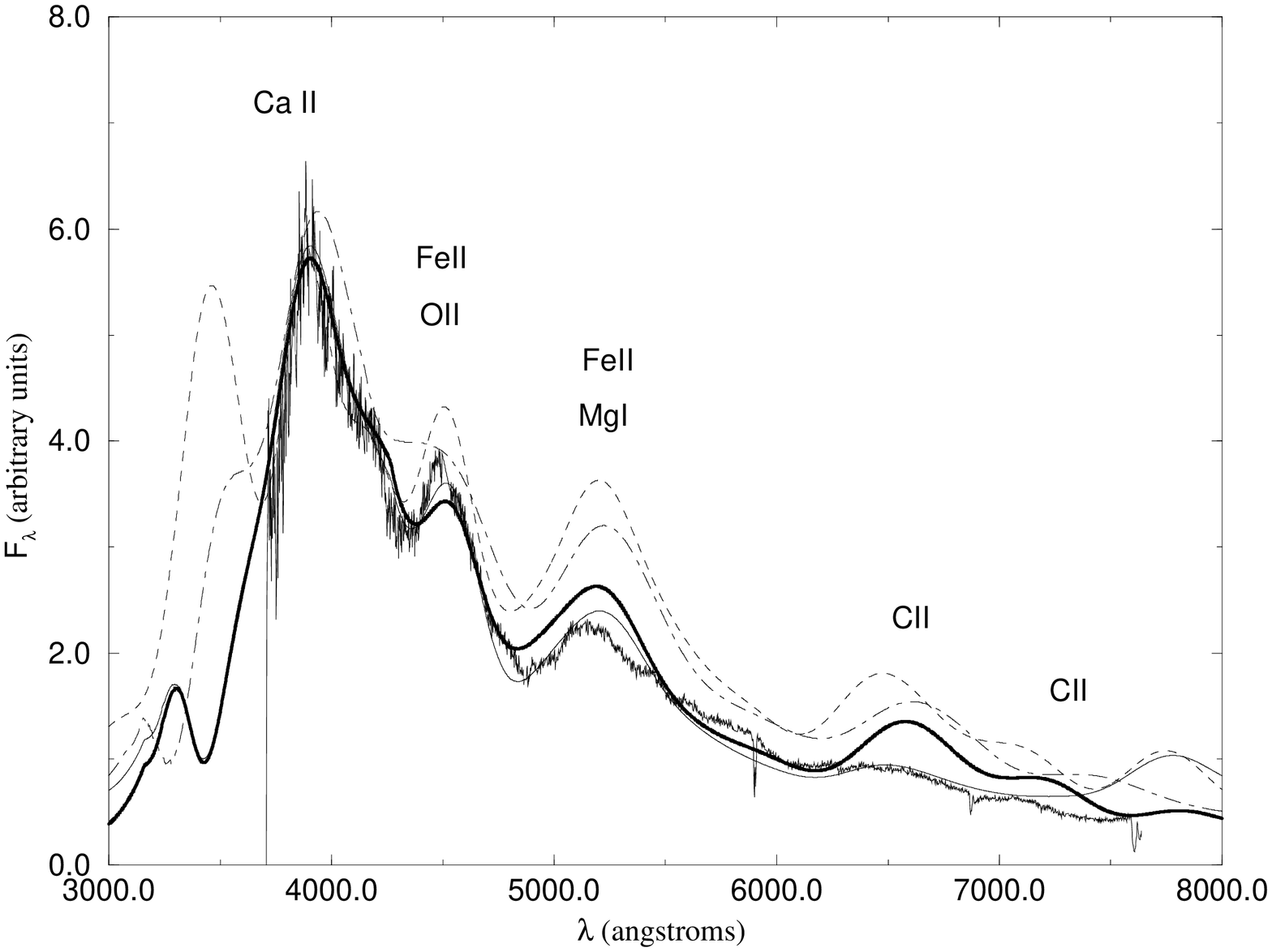,angle=270,width=0.9\textwidth}
\caption{\label{apr03_1886+1052+1730+1959}}Observed and synthetic
spectra for April 3, 
1994. The observed spectrum has been dereddened with E(B-V) =
0.45~mag. The absorption features near 6860~\AA\ and 7600~\AA\ are
telluric. For the calculated models the parameters are \Teff = 8300~K,
$\vno = 12000 $~\kmps, $N=6$, $\xni = 0.01$, \heoh = 12, \heoo = 0.3,
\coo = 0.5, and $z=3$ (thick solid line), the same parameters but
\heoo = 3.0 (dot-dashed line) and the same parameters but $N=8$, $\vno
= 15000$~\kmps, and \heoo = 0.3 (dashed line). The thin solid line
corresponds to a model with the same parameters as the thick solid
line, but C~II lines have been removed from the line-list in the
theoretical spectrum.
\end{figure*}

Figure~\ref{apr03_1886+1052+1730+1959} displays the data obtained on April
3.5, 1994 UT at the MMT as well as our calculated model with \Teff =
8300~K, $\vno = 12000 $~\kmps, $N=6$, $\xni = 0.01$, \heoh = 12, \heoo
= 0.3, \coo = 0.5, and $z=3$. Our boundary conditions were for the
outer radius and density were $R_{\rm out} = 1.96 \times 10^{15}$~cm
and $\rho_{\rm out} = 3.93 \times 10^{-17}$~\gcm. The line
identifications correspond to the results of our theoretical models
and we try to indicate where model limitations may be inaccurate.  The
synthetic spectrum is dominated by Ca~II and C~II lines,
although Fe~II and O~II lines are also present. The
synthetic spectrum fits the Ca~II H+K emission profile very
well, but the emission near 4500~\AA\ is too weak. It fits the broad
absorption feature near 5000~\AA\ reasonably well, but there is too
little absorption and the emission is too strong. In our model
spectrum the emission is due to Fe~II and Mg~I lines.  It
is likely that NLTE treatments of S~II, Si~II and
C~II would better reproduce the observed spectrum. Also plotted
in Figure~\ref{apr03_1886+1052+1730+1959} is the synthetic spectrum for a
model with $N=8$, giving a steeper atmosphere. The other parameters
were the same but in order to better reproduce the
observed spectrum $\vno = 15000$~\kmps.
 The lines are too narrow in the steeper
model in spite of the fact that the velocity is larger. As to be
expected, the steeper model only enhances the mismatch with the
observed spectrum in the red since the Fe~II and C~II
lines become more pronounced.  There is no clear evidence of
He~I lines in our synthetic spectrum and reducing the helium
abundance by a factor of 10 does not significantly affect the
spectrum. Increasing the helium abundance by a factor of 10 still
fails to show clear evidence for He~I lines, but the quality of
the overall fit is significantly reduced, (see
Figure~\ref{apr03_1886+1052+1730+1959}); the Ca~II H+K emission
becomes too broad, and the feature near 4500~\AA\ becomes very flat in
absorption with only weak emission.  Thus, at present, we can not
easily determine the amount of helium in the ejecta. Higher amounts of
helium do begin to show the effects of He~I. In the model with
high helium abundance (\heoo = 3.0), the helium mass fraction is
$Y=0.35$ and the total mass above the $\tau_{\rm std} = 1\ (30)$ is
$0.08\ (0.34)$~\msol, respectively.  For the model with \heoo = 0.3,
the helium mass fraction is $Y= 0.05$.  Near-IR spectra including the
He~I $\lambda10830$ line would be extremely useful for
determining the helium abundance.

\begin{figure*}
\psfig{file=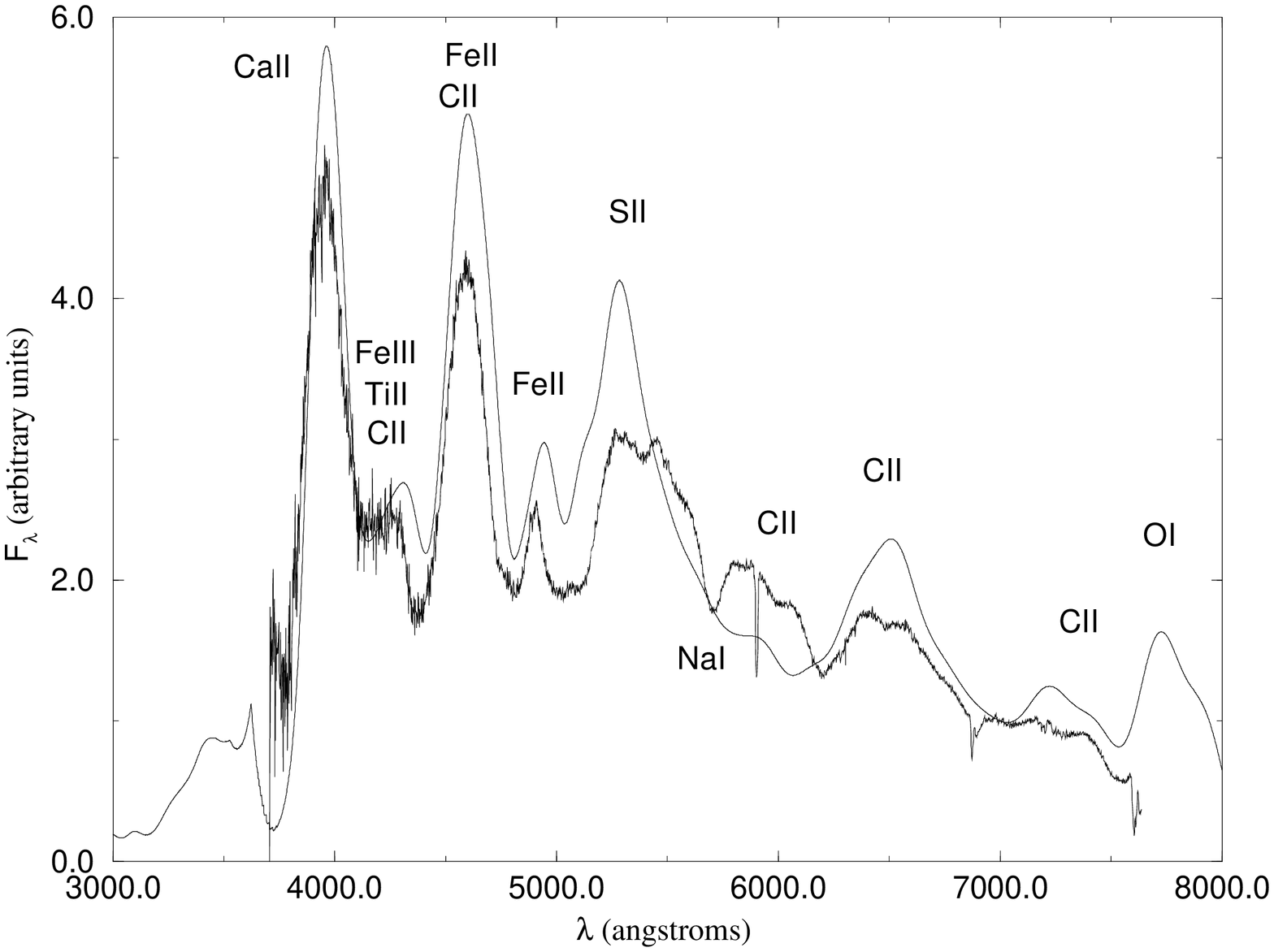,angle=270,width=0.9\textwidth}
\caption{\label{apr10_2065}}Observed and synthetic spectra for April 11.5,
1994. The observed spectrum has been dereddened with E(B-V) =
0.45~mag. The absorption features near 6860~\AA\ and 7600~\AA\ are
telluric.
\end{figure*}

We display in Figure~\ref{apr10_2065} the observed spectrum for April
11.5, 1994 UT obtained at the MMT along with our best fit synthetic
spectrum.  The model parameters are: \Teff = 6650~K, $\vno = 8000
$~\kmps, $N=8$, $\xni = 0.02$, \heoh = 12, \heoo = 0.001, \coo = 0.5,
$z=4$, $R_{\rm out} = 2.13 \times 10^{15}$~cm, and $\rho_{\rm out} =
5.04 \times 10^{-17}$~\gcm. The observed spectrum now displays lines
of S~II, Si~II, Na~I, and O~I, in addition
to Ca~II and Fe~II lines. The  O~I $\lambda 7773$ line is
clearly evident (and not too badly fit). The calculated Ca~II H+K
red absorption edge does not quite extend far enough to the blue
although the width of emission profile is about right.  The S~II
absorption to the red of 5000~\AA\ is not broad enough.  The observed
absorption feature near 5600~\AA, almost certainly due to Na~I,
is poorly fit. Also poorly fit is the observed absorption near
6200\AA, likely due to Si~II~$\lambda 6355$. A NLTE treatment of
Si~II and S~II would likely improve the agreement between
theory and observation. The high metallicity is required in order to
obtain good agreement with the spectra, and is consistent with the
position of the supernova close to the nucleus and observational
determinations of the metallicity of M51 \cite[]{diaz91}.

\begin{figure*}
\psfig{file=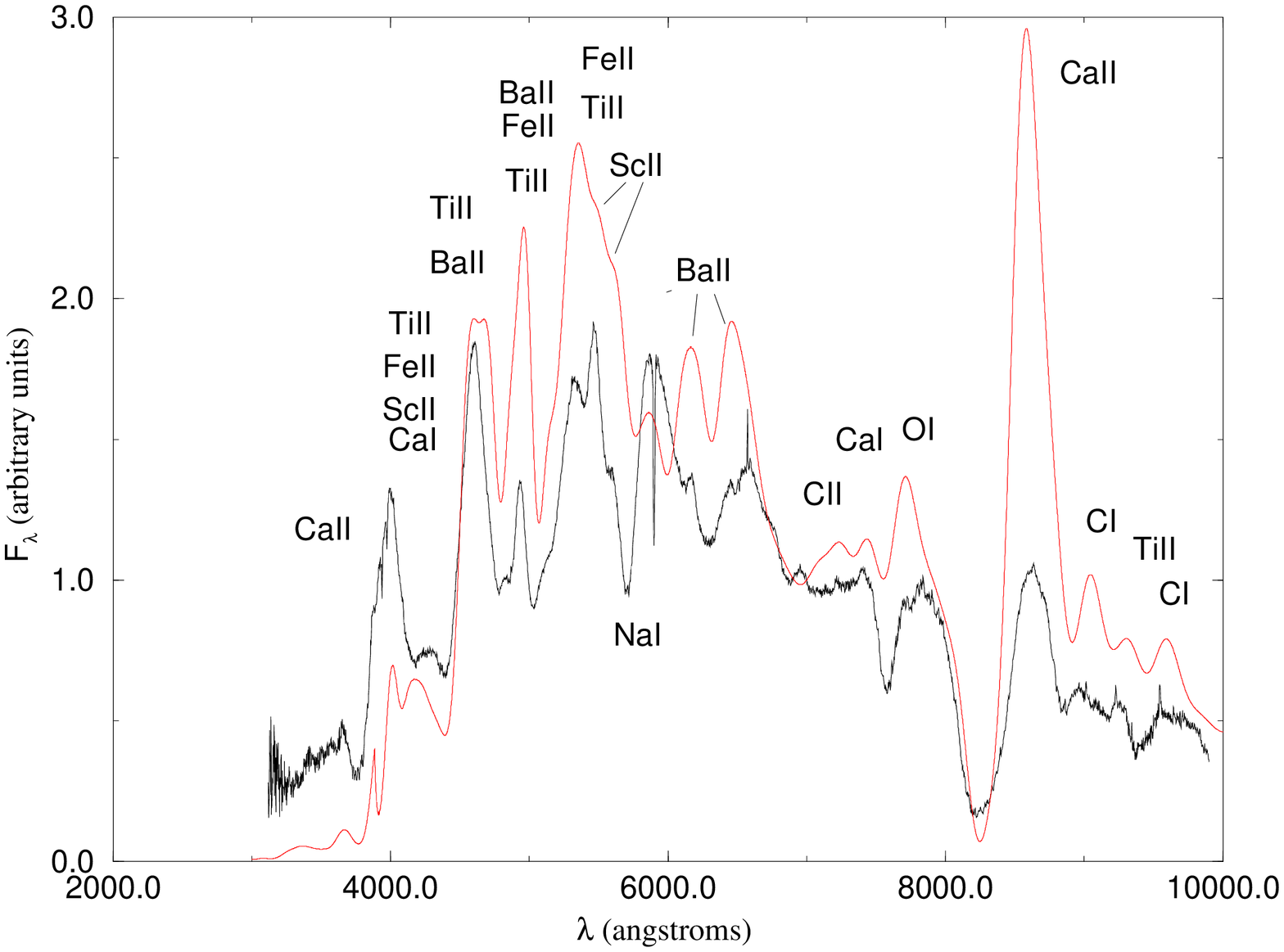,angle=270,width=0.9\textwidth}
\caption{\label{apr18_345}}Observed and synthetic spectra for April 18,
1994.The observed spectrum has been dereddened with E(B-V) =
0.45~mag. All telluric features have been removed.
\end{figure*}

The observed spectrum obtained on April 18, 1994 at Lick Observatory
is displayed in Figure~\ref{apr18_345}, along with our best fit
synthetic spectrum with model parameters: \Teff = 5300~K, $\vno = 8000
$~\kmps, $N=8$, $\xni = 0.01$, \heoh = 13, \heoo = 0.001, \coo = 0.5,
$z=4$, $R_{\rm out} = 3.9 \times 10^{15}$~cm, and $\rho_{\rm out} =
6.4 \times 10^{-17}$~\gcm.  The C~I $\lambda 9088$ and $\lambda 9641$
lines are clearly identified as is the O~I $\lambda 7773$ line, even
if the agreement with observations is poorer than desired. At this
time, the calcium profiles for the H+K lines and the IR triplet are
sensitive to the amount of \nni\ as well as to the density
profile. More nickel leads to narrower profiles since it ionizes
calcium at high velocity. The red edge of the calcium IR triplet
absorption profile is a bit too fast which may be due to the density
profile deviating from our assumed power-law. The red edge of the
calcium IR triplet can be better fit and the large emission in the
infrared triplet can be suppressed by making the atmosphere shallower,
but then the overall fit to the rest of the observed spectrum is
significantly degraded. We also find that better fits to the calcium
profile can be obtained with a lower velocity $\vno = 7000$~\kmps, but
this leads to large emission in the 4000--6000~\AA\ range that no
choice of temperature can improve, as well as a much poorer fit to the
O~I absorption.

\section{Reddening\label{redden}}

\begin{figure*}
\psfig{file=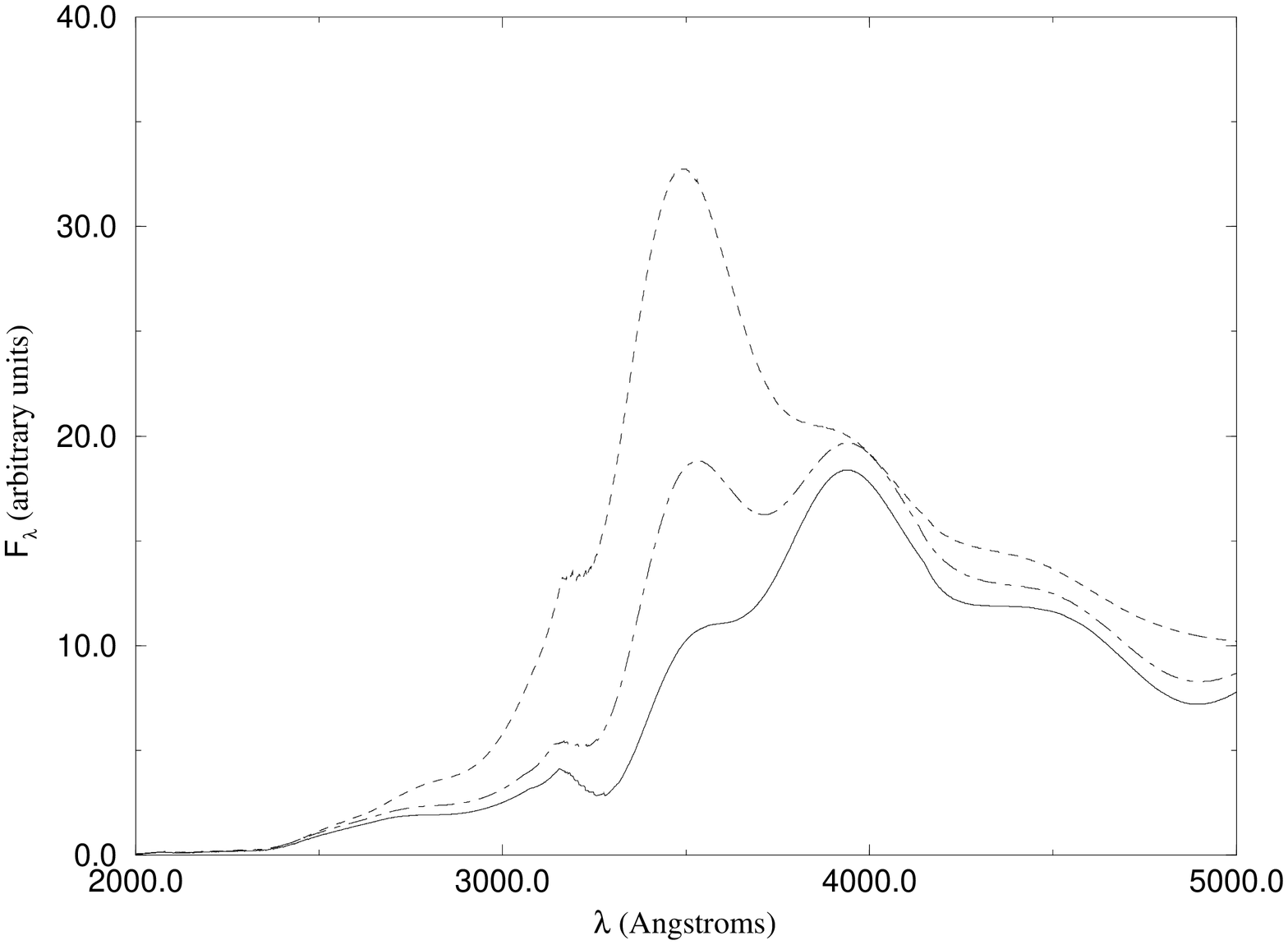,angle=270,width=0.9\textwidth}
\caption{\label{temps}}Synthetic spectra with \Teff = 8300 (solid
line), 8600 (dot-dashed line), and 9100~K (dashed line).
\end{figure*}

The strong interstellar sodium line visible in all observed spectra is
evidence for a significant amount of reddening in SN 1994I; 
The observed position of the supernova coincides with a dust finger in the
host galaxy M51. Based on an overall fit to the light curve Iwamoto
\etal\ (1995) find an extinction of A$_V=1.4$~mag, while an
analysis of the Na~D absorption 
 finds \({\rm
A}_V=3.1\stackrel{+3.1}{\scriptstyle -1.5}
\)~mag \cite[]{hf95}.  One can use the
earliest spectrum on April 3 to estimate the reddening. The observed
spectrum displays strong absorption in Ca~II H+K. However, as
the reddening increases, the color temperature of the synthetic
continuum spectrum must also increase in order to match the observed
spectrum.  But this higher temperature depopulates the lower level of
the Ca H+K lines and hence the strong absorption seen in the spectrum
cannot be reproduced. Figure~\ref{temps} displays our calculated
spectra for \Teff = 8300, 8600, and 9100~K. We find that the
temperature is fairly well determined $\Teff \approx 8300$~K, although
it may be possible to go to higher temperatures by confining calcium
to cooler regions. On the other hand, our models produce reasonably
good agreement with the red edge of the calcium absorption which
indicates that we do not have calcium mixed too deeply.  Using the
reddening law of \cite{card89} we
find that extinctions $0.9 \la A_V \la 1.4$~mag
 provide good fits to the
observed spectrum at this epoch, and  that $A_V \approx 1.4$~mag is
clearly favored at later epochs.

\section{Discussion and Conclusions}

The nickel abundance is not well determined by our models due to the fact
that our deposition function is local, and even at maximum light  not
all of the gamma rays are trapped with
such low mass ejecta. 
Our results are consistent with the small nickel masses
implied by light curve analyses.
	
\cite{whsn94i95} have identified the
absorption feature near 6200~\AA\ as being due to
Si~II~$\lambda6355$. Although in all of our models the absorption near
6200~\AA\ is due to C~II and Ba~II, the fit to observations is poor.
In order to correctly fit the observed spectrum, it is likely that
Si~II needs to be treated in NLTE.  It is not surprising that our
results are different from those of Wheeler \etal\ (1995); even though
both calculations treat Si~II in LTE, our self-consistent temperature
structure is likely to differ from the grey temperature structure of
Wheeler \etal\ (1995).

The uncertain rise-time and reddening of SN 1994I make it a less than
ideal candidate for the spectral-fitting expanding atmosphere method
\cite[SEAM, see][]{b93j3,nug1a95}; note that we have corrected an
error in our synthetic photometry that was present in those
papers]. Nevertheless, the absolute flux is a direct result of our
models, and the predicted distances serve as a useful check on the
accuracy of the synthetic spectra.  Applying this method to SN 1994I,
with an assumed reddening of $E(B-V) = 0.45$~mag, we find a bolometric
rise-time of 9 days, and a distance modulus to the supernova of $\mu =
28.9 + 5\log(t_r/9\ {\rm days}) \pm 0.5 \pm 0.7$~mag, where the first
uncertainty is simply the standard deviation in all our fits and the
second uncertainty is an estimate of our systematic error
\cite[]{nug1a95} including an estimate of the error due to the
uncertain reddening correction. In principle, the distance estimate
gives a value for the rise-time, by minimizing the scatter in distance
obtained at different epochs, but for SN 1994I the uncertain
reddening, and the use of only 3 epochs with one before maximum light,
makes such a determination ambiguous.  The rise-time should which
should be compared to the results of \cite{iwsn94i94}, who find a 12
day rise-time for models of the light-curve of SN 1994I.

Comparing our results to published hydrodynamical models of the light
curve of SN~1994I, we find that our models have substantial mass above
the reference radius.  On April 18, the mass above the reference
radius ranges from 0.5~\msol\ for our shallowest models to 1.6~\msol\
for our steepest models. The steeper, more massive model produces a
significantly better fit to the observed spectrum than the shallower,
less massive model, but since the Ca~II profile is sensitive to the
density profile and given our simplifying assumptions of uniform
compositions and a power law density profile it would be premature to
draw conclusions about the density structure from our models.

Although the two published hydrodynamical models of the light curve of
SN 1994I are from very different stellar evolution scenarios
\cite[]{wlw95,iwsn94i94}, the actual models are very similar in total
ejected mass, total nickel mass, and energy of explosion. Without
performing detailed synthetic spectral modeling of each calculation
it is difficult for us to favor one stellar evolution scenario or the
other.

Our calculations lend credence to the identification of SNe Ic
progenitors as C+O cores formed in an interacting binary
\cite[]{wh90}. In order to make definitive statements about
abundances, more intermediate mass elements need to be included in
NLTE. Additionally, non-uniform compositions need to be
treated. However, once the atmospheres become inhomogeneous, parameter
space becomes larger than can be adequately covered. A better way to
proceed is to perform spectroscopic diagnostics on hydrodynamical
models. In order to do this in an objective manner, hydrodynamical
calculations must fully resolve the density profile in the atmosphere.

\section*{Acknowledgments}
We are most grateful to Brian Schmidt for providing the MMT spectra
quickly and for helpful comments on the manuscript.  We thank Adam
Fisher and Sumner Starrfield for helpful discussions, and Ken Nomoto
for providing us with the results of his hydro models.  Assistance
with the observations and reductions at Lick Observatory was provided
by Luis C. Ho and Aaron J. Barth.  This work was supported in part by NASA
grant NAGW-2999, a NASA LTSA grant to ASU, NASA grant GO-2563.01-87A
from the Space Telescope Science Institute, which is operated by the
Association of Universities for Research in Astronomy, Inc., and by
NSF grants AST-9115061 and AST-9115174.  Some of the calculations in
this paper were performed at the NERSC, supported by the U.S. DoE, and
at the San Diego Supercomputer Center, supported by the NSF; we thank
them for a generous allocation of computer time.


\vfill\eject

\end{document}